\documentclass[iop]{emulateapj}
\begin{document}

\title{The ground-based H, K, and L-band absolute emission spectra of HD~209458b}

\author{Robert T. Zellem$^{1}$, Caitlin A. Griffith$^{1}$, Pieter Deroo$^{2}$, Mark R. Swain$^{2}$, Ingo P. Waldmann$^{3}$}

\affil{$^{1}$ Lunar and Planetary Laboratory, University of Arizona, 1629 E. University Blvd., Tucson, AZ 85721, USA}
\affil{$^{2}$ Jet Propulsion Laboratory, California Institute of Technology, 4800 Oak Grove Drive, Pasadena, CA 91109, USA}
\affil{$^{3}$ University College London, Department of Physics \& Astronomy, Gower Street, London WC1E 6BT, UK}

\email{rzellem@lpl.arizona.edu}

\begin{abstract}

Here we explore the capabilities of NASA's 3.0~meter Infrared Telescope Facility (IRTF) and SpeX spectrometer and the 5.08~meter Hale telescope with the TripleSpec spectrometer with near-infrared H, K, and L-band measurements of HD~209458b's secondary eclipse. Our IRTF/SpeX data are the first absolute L-band spectroscopic emission measurements of any exoplanet other than the hot Jupiter HD~189733b. Previous measurements of HD~189733b's L-band indicate bright emission hypothesized to result from non-LTE CH$_{4}$ $\nu_{3}$ fluorescence. We do not detect a similar bright 3.3~$\micron$ feature to $\sim$3$\sigma$, suggesting that fluorescence does not need to be invoked to explain HD~209458b's L-band measurements. The validity of our observation and reduction techniques, which decrease the flux variance by up to 2.8 orders of magnitude, is reinforced by 1$\sigma$ agreement with existent $Hubble$/NICMOS and $Spitzer$/IRAC1 observations which overlap the H, K, and L-bands, suggesting that both IRTF/SpeX and Palomar/TripleSpec can measure an exoplanet's emission with high precision.

\end{abstract}

\section{Introduction}

Space-based telescopes offer the ability to make high-precision measurements of transiting exoplanets {\citep[e.g.,][]{knutson08, swain08, crossfield12b, zellem14, diamondlowe14}} However, these platforms have limited time and instrumentation compared to ground-based capabilities. Yet, ground-based platforms are plagued by atmospheric attenuation and systematic errors which increase a dataset's variance, making it difficult to extract an exoplanetary signal.

Here we investigate the potential of ground-based measurements with NASA's 3.0 meter Infrared Telescope Facility (IRTF) with the SpeX spectrometer and the 5.08 meter Hale telescope with the TripleSpec spectrometer at Palomar Observatory. Our observations  of the bright exoplanetary system HD~209458b span the infrared H, K, and L-bands and include the first transiting exoplanet observations with Palomar/TripleSpec. These wavelengths have also been been exposed by the $Hubble$ and $Spitzer$ space telescopes, allowing us to test our own methodologies for treating the extraction of the planetary signal from atmospheric effects and systematic errors. Our ultimate goal is demonstrating the capability of both platforms for near-IR transiting exoplanet spectroscopy.

This project also includes the first absolute L-band ($\sim$3--4~$\micron$) spectroscopic measurements of an exoplanet other than the hot Jupiter HD~189733b. The 4 spectra recorded of HD~189733b reveal a bright (T$_{bright}$ $\approx$ 2750~K, compared with HD~189733b's T$_{eff} \approx 1200$~K) emission source at 3.3 $\micron$ with a wavelength and shape that are consistent with the P, Q, and R branches of the CH$_{4}$ $\nu$$_{3}$ vibrational-rotational band \citep{swain10,waldmann12}. This feature is also observed in the spectra of Titan \citep{kim00}, Jupiter \citep{drossart99, brown03}, and Saturn \citep{drossart99}, where its large flux is explained by fluorescence. Since HD~209458b (T$_{eff} \approx 1450$~K) is hotter than HD~189733b, it has a comparably lower CH$_{4}$ content, according to thermochemical equilibrium \citep{moses11}, and potentially no bright L-band feature. We search here for a similar feature, or lack thereof, in HD~209458b's L-band spectrum to further investigate the radiative mechanisms of exoplanetary atmospheres.

\section{Observations and Data Reduction}
We observed HD~209458b's emission with the 3.0 meter NASA Infrared Telescope Facility (IRTF) at Mauna Kea Observatory and SpeX \citep{rayner03}, a near-IR spectrometer with a wavelength coverage of 2.0--4.2 $\micron$ (K and L-bands) and a resolution of R = 2500, and with the 200 inch (5.08~meter) Hale Telescope at Palomar Observatory and TripleSpec, a near-IR spectrometer with a wavelength coverage of 1.0--2.4 $\micron$ (J, H, and K-bands) and a resolution of R = 2500--2700. While low-resolution spectroscopic observations are incapable of observing the fine scale structure of the spectral lines, the SpeX and TripleSpec spectral channels can be binned to increase the signal-to-noise ratio (SNR). Thus IRTF/SpeX and Palomar/TripleSpec can observe the broad spectral behavior of an exoplanet such as HD~209458b, allowing them to complement their high-resolution counterparts (e.g., the R = 100,000 VLT/CRIRES). The capability of IRTF/SpeX for exoplanet spectroscopic observations was verified by the observations of \cite{swain10}, who binned spectral bands unaffected by terrestrial absorption in Fourier-space to reduce ground-based systematic errors and reproduce the \textit{Hubble} K-band spectra of HD~189733b. Their results were confirmed by \citet{thatte10} using a method involving a principal component analysis (PCA) and reproduced by \citet{waldmann12}, despite the variable humidity between their four nights of observations. Others \citep[e.g.,][]{richardson03b, crossfield12, danielski14} have also used IRTF/SpeX for additional eclipsing exoplanet observations. The capability of Palomar/TripleSpec for transiting exoplanet spectroscopy has yet to be demonstrated. The wavelengths covered by SpeX and TripleSpec are also partly covered by previous \textit{Hubble}/NICMOS \citep{swain09a} {and $Spitzer$/IRAC 1 \citep{knutson08, diamondlowe14} measurements} of HD~209458b's emission, thereby providing important verification for our difficult ground-based observations.


\subsection{IRTF/SpeX} \label{section:irtf}
We observed HD~209458b's 2011 September 9 (UT) secondary eclipse for $\sim$8 hours, resulting in 1210 exposures of 10 seconds each in an ABBA nodding sequence. Systematic errors can be introduced by telescope jitter caused by, for example, slewing to a comparison or check star or adjusting the telescope's focus. Therefore, to limit the introduction of additional systematics, we set the focus only once at the beginning of the night and stayed on the target throughout the entire night. Error from seeing- and pointing-induced signal loss is minimized by a 1.6'' wide slit; the seeing was at best 0.47'' and at worst 1.51''. The raw images are dark-subtracted, flat-fielded, and wavelength-calibrated. The spectrum is extracted from each image using the Spextool reduction package for IDL \citep{cushing04, vacca03}.



\begin{figure}
\includegraphics[width=1\columnwidth]{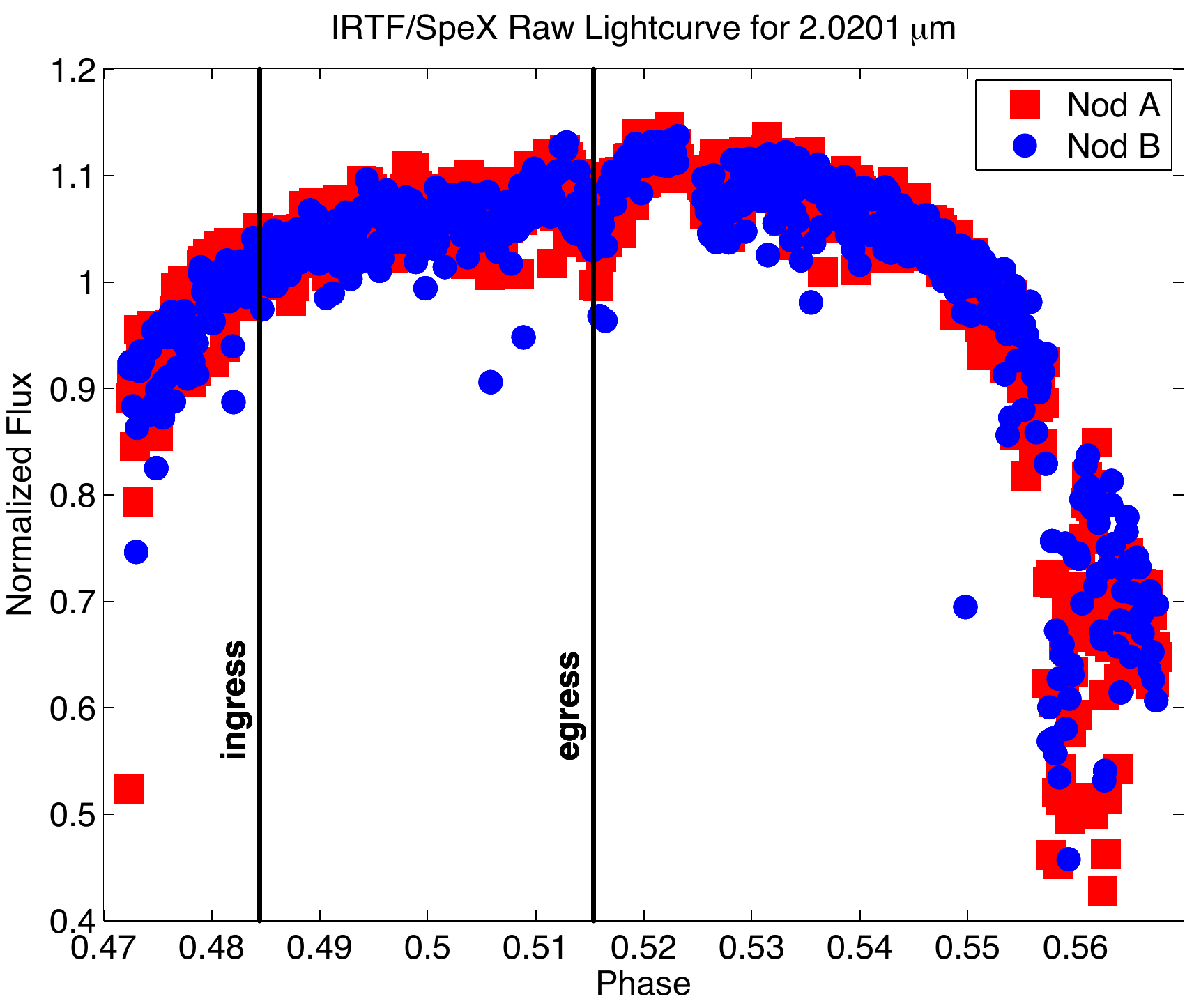}
\caption{A sample IRTF/SpeX raw lightcurve extracted from one wavelength channel (2.0201 $\micron$) of 900 total wavelength channels. At this point, systematic errors cause a variance of $\sim$10 \%, preventing the detection of HD~209458b's smaller $\le$ 0.2 \% eclipse signal (horizontal lines indicate the predicted ingress and egress eclipse times). Nod A (red squares) and nod B (blue circles) are also plagued by their own distinct set of systematic errors. Therefore we run each nod separately through the reduction and combined them at the very end.}
\label{fig:rawlc_irtf}
\end{figure}

\begin{figure}
\epsscale{1}
\includegraphics[width=1\columnwidth]{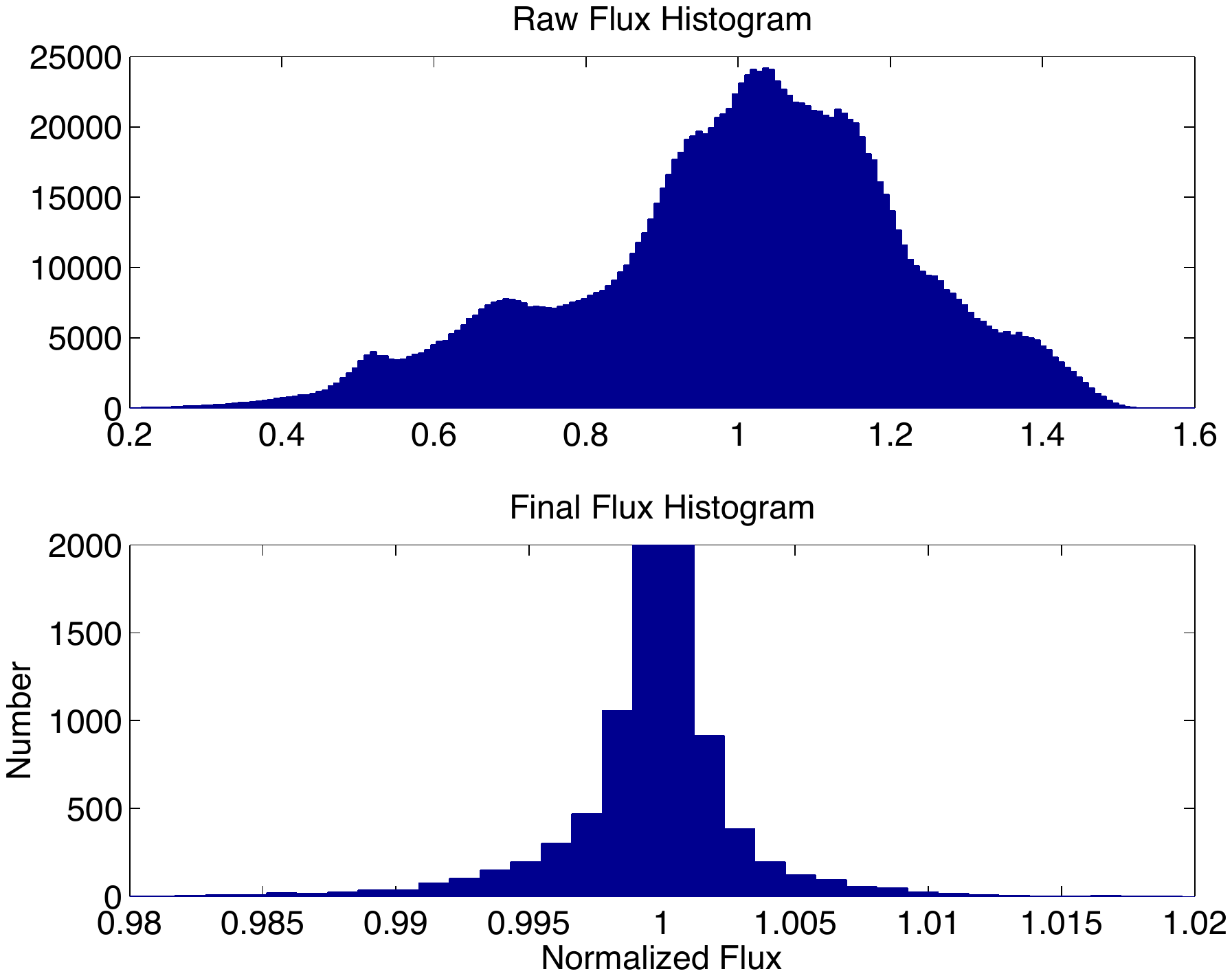}
\caption{\textit{Top:} Histogram of the nod A normalized raw K and L-band IRTF/SpeX flux. Note not only the large variance but also the non-symmetrical distribution of the fluxes. \textit{Bottom:} Histogram of the normalized K and L-band flux at the end of the entire reduction for nod A. A normal symmetrical distribution is restored and the overall variance is reduced by 1.6 orders of magnitude.}
\label{fig:fluxhist_irtf}
\end{figure}

For ground-based observations, typical sources of error include changing airmass as the target rises and sets, telescope jitter, or mirror flexure. Such errors can cause data scatter on the order of $\sim$10--20\% (but sometimes even as large as $\sim$40--60\%), which dwarf HD~209458b's expected secondary eclipse depth of $\le$0.2\% (Figure~\ref{fig:rawlc_irtf}). These non-Gaussian error sources are reflected in the raw flux histogram (Figure~\ref{fig:fluxhist_irtf} \textit{Top}) and must be removed or reduced to detect the small planetary signal. To minimize error propagation from telluric absorption lines, we select wavelength channels that are not close to the band edges: 2--2.38 $\micron$, corresponding to 900 spectral channels in the K-band, and 3--3.99 $\micron$, corresponding to 1400 spectral channels in the L-band. All observations at comparably higher airmass values ($\ge$ 1.65; corresponding to a phase $\ge$ 0.55 in Figure~\ref{fig:rawlc_irtf}) and just after the telescope flipped the pier (0.5175 $\le$ phase $\le$ 0.5215) were also removed.

As with previous IRTF/SpeX observations \citep{swain10}, the time-varying flux (Figure~\ref{fig:rawlc_irtf}) indicates that the A and B nods are differently affected by systematic errors (e.g., due to varying pixel efficiencies on different parts of the detector). Therefore we separate the two nods, reduce each set independently to eliminate nod-specific systematics, and then merge them at the end. Anomalies, such as cosmic rays, are removed with a two-dimensional Wiener filter. This filter estimates the mean and variance for each pixel from its 8 nearest neighbors and replaces an outlying pixel with the mean of its surrounding pixels.

As discussed in detail below, our reduction follows these steps: [1] normalize the data to reduce the variance due to systematic errors, [2] fit the time-varying flux to an exponential curve to correct for airmass, [3] remove outliers, [4] enhance the eclipse signal and reduce the signal due to noise by binning data in Fourier space, and [5] flatten residual curvature.


This reduction is based upon and uses the same assumptions of the ``self-coherence'' methods of \citet{swain10} and \citet{waldmann12}, which assume that (1) the systematic errors are large with respect to signal, (2) the systematic errors are based either on correlations in wavelength or time, and (3) the difference between eclipse depths in adjacent wavelength channels is small.


Systematics can affect each spectrum differently, causing individual spectra to have unique brightness variations which manifest as intra-spectral variations. To reduce these variations, we divide the data into blocks of 100 spectral channels and then normalize each spectrum's time-varying flux by the mean of its bin's time-varying flux:
\begin{eqnarray}
F_{i}^{j}=\frac{S_{i}^{j}}{S_{avg,i}} \label{eqn:normalize}
\end{eqnarray}
where
\begin{eqnarray}
S_{avg,i}=\frac{\sum_{j=1}^{n} S_{i}^{j}}{n} \label{eqn:mean}
\end{eqnarray}
and $\textit{i}$ is the time index, $\textit{j}$ is the wavelength channel, and $\textit{n}$ is the number of wavelength channels in the block (\textit{n} = 100). This normalization divides out wavelength-dependent and time-dependent errors common among the wavelength channels in each bin \citep[referred to as ``common-mode'' errors in][]{swain10}. Since the eclipse signal is too small to be detected at this point, only systematic errors are removed and not the commonly-shared eclipse signal.


At this point, airmass variations dominate the data. As seen in Figure ~\ref{fig:rawlc_irtf}, the data have a different shape before and after HD~209458 reaches the zenith. This phenomenon is due to the telescope settling throughout the beginning of the night. Therefore each half of the lightcurve (pre- and post-zenith) is fit with an exponential airmass correction:
\begin{eqnarray}
a^{j} \cdot e^{b^{j} \cdot airmass_{i}} = F_{i}^{j}
\end{eqnarray}
where the airmass is a function of time $\textit{i}$ and \textit{j} is the wavelength channel index, such that each half of the time series (for each wavelength channel) is individually corrected via the equation:
\begin{eqnarray}
I_{i}^{j} = \frac{F_{i}^{j}}{a^{j} \cdot e^{b^{j} \cdot airmass_{i}}}
\end{eqnarray}
where $I_{i}^{j}$ is each individual airmass-corrected channel $j$ at time $i$.


While this correction flattens the lightcurves, some outliers persist. We statistically identify these outliers with Chauvenet's criterion: if a datapoint from a population of \textit{n} datapoints has a Gaussian probability less than $\frac{1}{2\textit{n}}$, then this datapoint is an outlier. We use Chauvenet's criterion in a running boxcar with a width \textit{n} = 10 on each lightcurve's time-varying flux and replace any outliers with the mean of the \textit{n} datapoints.

However at this point the eclipse signal is still too small to be detected.  This low frequency eclipse signal is enhanced by taking the geometric mean of each block in Fourier space \citep{swain10}:
\begin{eqnarray}
F_{i}^{k} = IFT\Bigg[\bigg(\prod_{j=1}^{n}FT\big(I_{i}^{j}\big)\bigg)^{\frac{1}{n}}\Bigg] \label{eqn:fft}
\end{eqnarray}
where \textit{I} is the channel's flux, $\textit{i}$ is the time index, $\textit{j}$ is the wavelength channel index, \textit{FT} is the Fourier transform, $\textit{n}$ is the number of channels per bin (\textit{n} = 100), $\textit{k}$ is the new bin index ($k = 1-9$ for the K-band; $k=10-23$ for the L-band), and \textit{IFT} is the inverse Fourier transform. This method converts the data from time-space to frequency-space then enhances the eclipse signal by taking the geometric mean of 100 spectral channels to form a new wavelength bin. All modes are kept and then this bin is then converted back from frequency- to time-space with the inverse Fourier transform $IFT$. The end result is 9 new K-band spectral bins, due to binning 900 spectral channels by 100, and 14 new L-band spectral bins, due to binning 1400 spectral channels by 100.


However some residual systematic errors (e.g., airmass and possibly some curvature introduced by the Fourier transform) remain, as indicated by residual curvature in the lightcurves. In order to correct for this curvature, we fit various amounts of baseline (out-of-eclipse points) to a second degree polynomial. Since the amount of pre-eclipse data is limited due to the eclipse ingress being comparatively close to sunset (Figure~\ref{fig:rawlc_irtf}), we vary the pre-eclipse baseline only up to half its length and allow the total number of baseline points to vary as long as their number is always greater than or equal to the number of in-eclipse (second to third contact) points. As a result, 3783 different amounts of baseline per lightcurve are fit to a second degree polynomial. The optimal amount of baseline is the one that best minimizes the out-of-eclipse baseline scatter. Its corresponding second-degree polynomial fit is then interpolated to the in-eclipse data to flatten the entire lightcurve. The eclipse depth is then extracted from the lightcurve by fitting a \citet{mandel02} model lightcurve using the physical parameters of R$_{star}$ = 1.146R$_{\sun}$ \citep{brown01}, R$_{planet}$ = 1.38R$_{Jupiter}$, orbital inclination of 86.59$\degr$, semimajor axis of 0.04747 AU \citep{southworth10}, orbital period of 3.524746 days \citep{torres08}, and period epoch of 2455216.405640 BJD\_UTC  \citep{zellem14}.

This baseline fitting routine can potentially introduce additional uncertainty to the derived eclipse depth. We estimate this error with a Monte Carlo simulation of the baseline fitting routine with 3.458 million lightcurves per original lightcurve, generating 3.458 million eclipse depths. The standard deviation of all of these simulated eclipse depths reflects the uncertainty of the final lightcurves' eclipse depths and, as a result, the calculated planet-to-star flux ratio.

\begin{figure}
\epsscale{1}
\includegraphics[width=1\columnwidth]{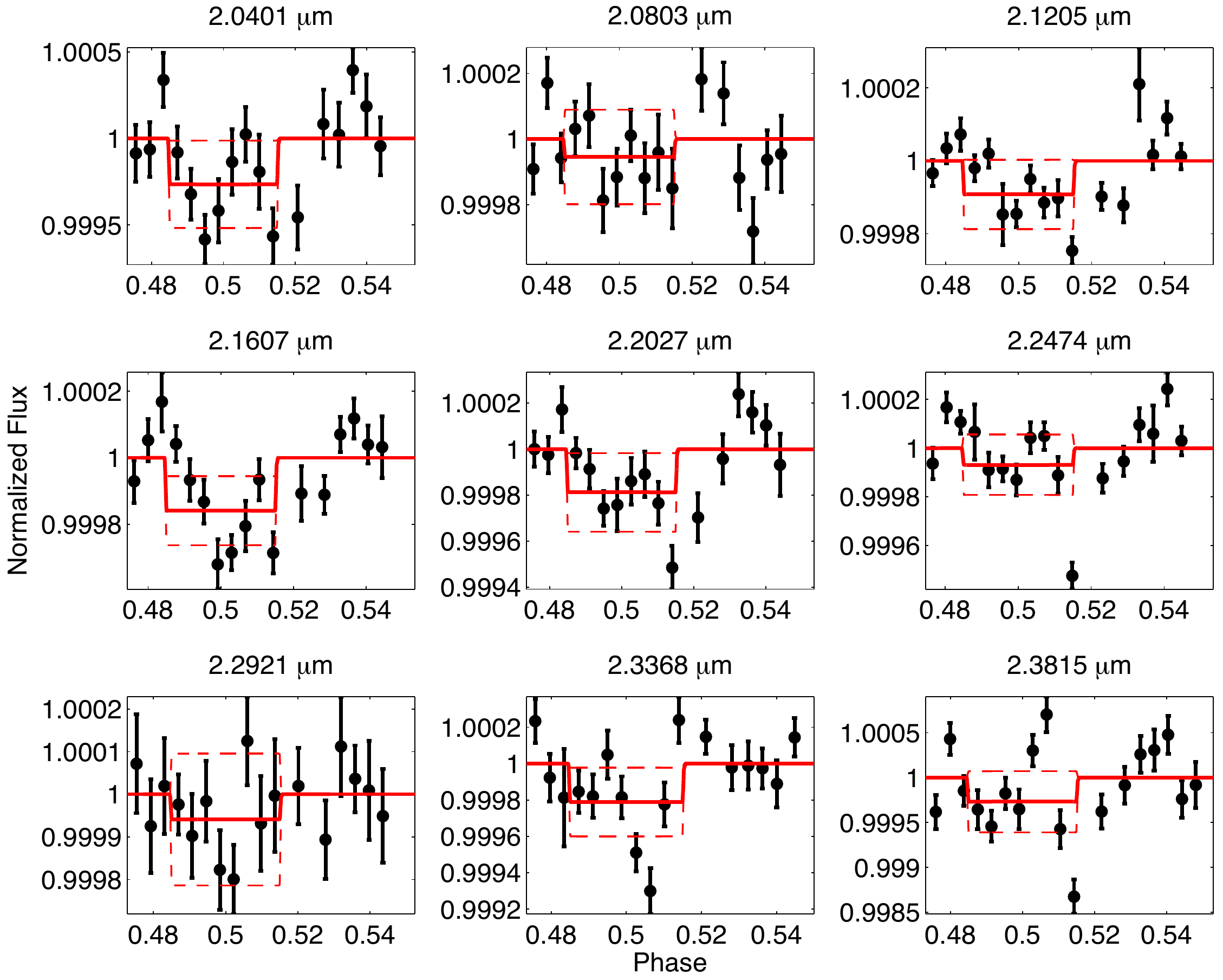}
\caption{The reduced IRTF/SpeX K-band lightcurves (both nods combined) with the data binned in time by 50 datapoints for clarity. The red line is a model lightcurve fit to the data. The dashed lines indicate the 1$\sigma$ uncertainty in our eclipse depth measurements.}
\label{fig:finallcs_irtf_k}
\end{figure}

\begin{figure}
\epsscale{1}
\includegraphics[width=1\columnwidth]{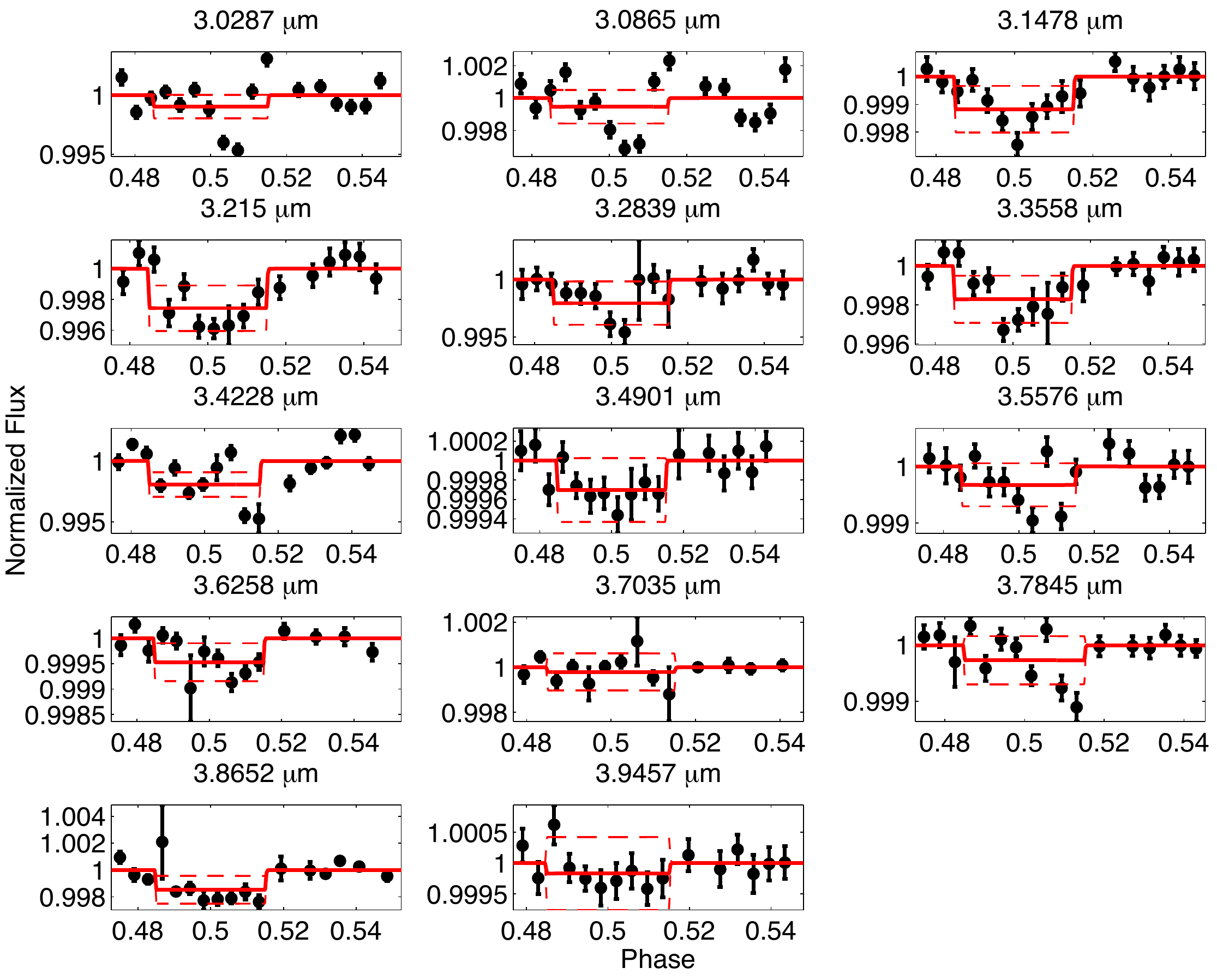}
\caption{The reduced IRTF/SpeX L-band lightcurves (both nods combined), following the protocol of Figure~\ref{fig:finallcs_irtf_k}.}
\label{fig:finallcs_irtf_l}
\end{figure}

\begin{figure}
\epsscale{1}
\includegraphics[width=1\columnwidth]{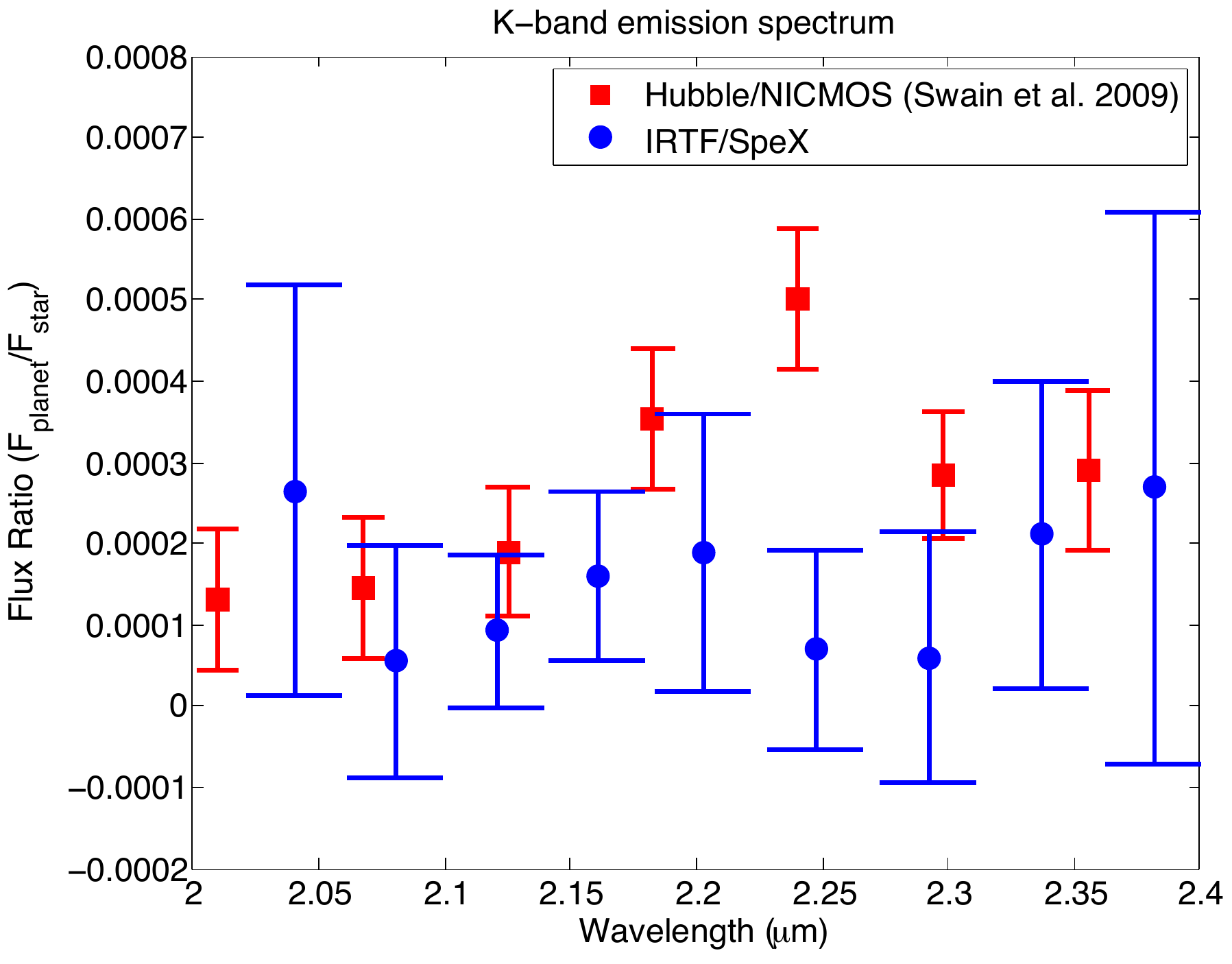}
\caption{Our final IRTF/SpeX K-band spectrum (blue circles) compared to the \citet{swain09a} \textit{Hubble}/NICMOS spectrum (red squares); please note that we use here a statically independent subset of the published \citet{swain09a} data. Our data agrees with the \textit{Hubble} data points, as indicated by the fact that nearly all of our 1$\sigma$ error bars overlap the \citet{swain09a} datapoints, reinforcing that IRTF/SpeX has great potential for eclipsing exoplanet spectroscopy.}
\label{fig:spectrum_irtf_k}
\end{figure}

\begin{figure}
\epsscale{1}
\includegraphics[width=1\columnwidth]{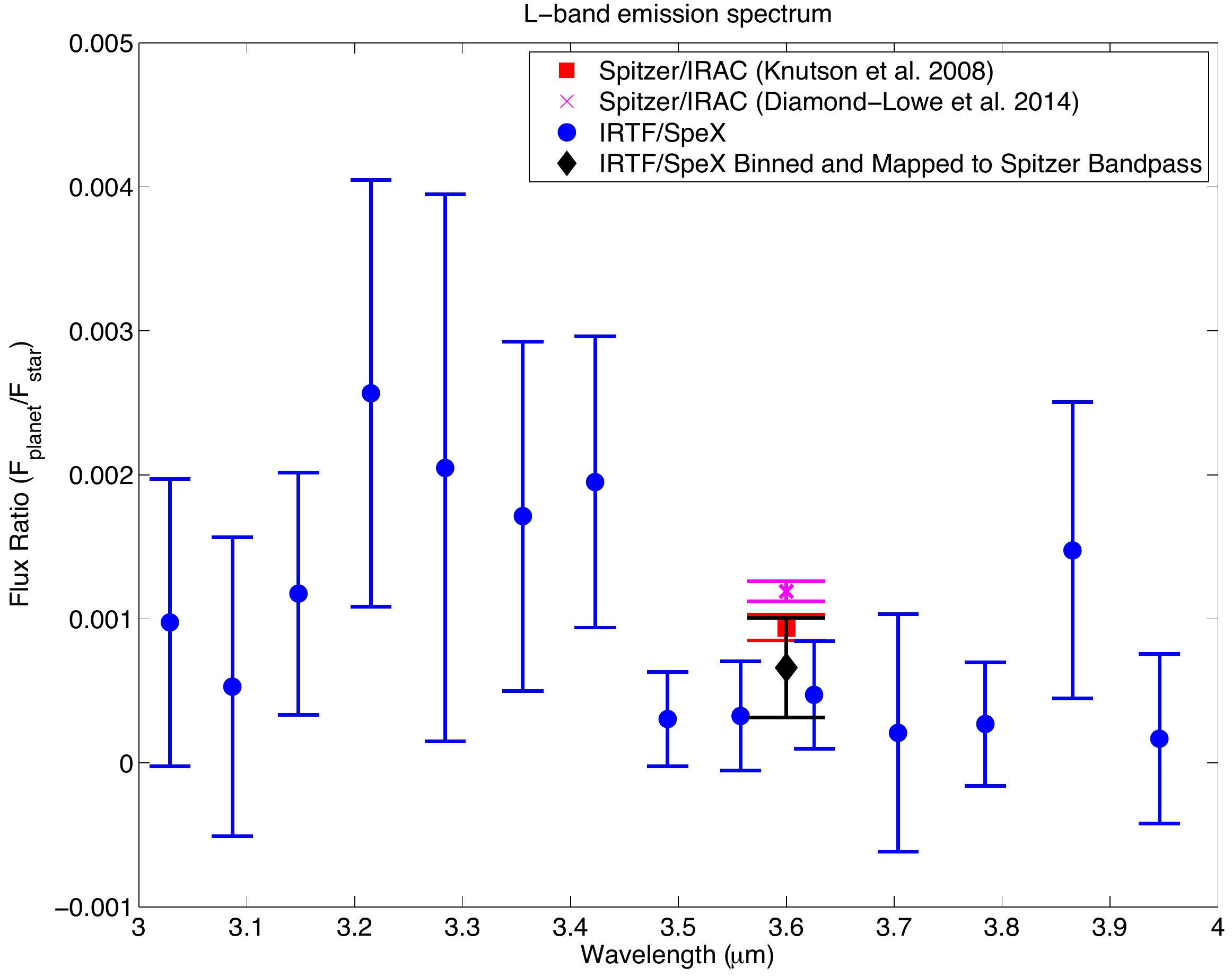}
\caption{Our final IRTF/SpeX L-band spectrum (blue circles) and binned to the $Spitzer$/IRAC 3.6 $\micron$ bandpass (black diamond). The binned IRTF/SpeX data agrees to 1$\sigma$ with a previous \citet{knutson08} $Spitzer$/IRAC1 emission measurement (red square) {and to 1.5$\sigma$ with a new $Spitzer$/IRAC1  emission measurement \citep{diamondlowe14} of HD~209458b (magenta cross)}, suggesting the validity of the IRTF/SpeX data.}
\label{fig:spectrum_irtf_l}
\end{figure}

The flux ratios of the planet to the star (F$_{planet}$/F$_{star}$) are calculated from the eclipse depths of the final reduced lightcurves (Figures~\ref{fig:finallcs_irtf_k} and~\ref{fig:finallcs_irtf_l}) and then used to construct the low resolution emission spectrum of HD~209458b (Figures~\ref{fig:spectrum_irtf_k} and~\ref{fig:spectrum_irtf_l}). The 1$\sigma$ errorbars for each spectral point are calculated from the root-sum-square of the standard deviations of the in- and out-of-eclipse data points as well as the error introduced from the baseline fit. The uncertainty of the eclipse depth, and therefore the planet-to-star flux ratio (F$_{planet}$/F$_{star}$), is due to not only errors introduced by the baseline fitting routine but also the scatter in the in- and out-of-eclipse data. Therefore, we conservatively estimate the 1$\sigma$ errorbars on our eclipse depths by adding in quadrature the uncertainties from the in-eclispe data, the out-of-eclipse data, and then the error introduced by the baseline fit.

\begin{deluxetable*}{c c c c c}
\tablewidth{0pt}
\tablecaption{IRTF/SpeX Measurements of HD~209458b's Emission}
\tablehead{
\colhead{Wavelength} & \colhead{F$_{\text{planet}}$/F$_{\text{star}}$} & \colhead{Uncertainty} & \colhead{Photon Noise} & \colhead{$\sigma/\sigma_{photon}$}\\
\colhead{($\micron$)} & \colhead{($\times$10$^{3}$)} & \colhead{$\sigma$ ($\times$10$^{3}$)} & \colhead{$\sigma_{photon}$ ($\times$10$^{3}$)} & \colhead{}}
\startdata

    2.0401&    0.2652&    0.2522&    0.0391&    6.4484\\
    2.0803&    0.0548&    0.1439&    0.0354&    4.0666\\
    2.1205&    0.0921&    0.0951&    0.0328&    2.9012\\
    2.1607&    0.1592&    0.1038&    0.0330&    3.1503\\
    2.2027&    0.1880&    0.1710&    0.0349&    4.8998\\
    2.2474&    0.0688&    0.1241&    0.0349&    3.5533\\
    2.2921&    0.0593&    0.1547&    0.0345&    4.4842\\
    2.3368&    0.2111&    0.1886&    0.0350&    5.3823\\
    2.3815&    0.2688&    0.3400&    0.0363&    9.3771\\
    3.0287&    0.9747&    0.9976&    0.0650&   15.3561\\
    3.0865&    0.5286&    1.0381&    0.0641&   16.1913\\
    3.1478&    1.1756&    0.8403&    0.0606&   13.8753\\
    3.2150&    2.5665&    1.4820&    0.0697&   21.2674\\
    3.2839&    2.0470&    1.8989&    0.0673&   28.2092\\
    3.3558&    1.7134&    1.2129&    0.0685&   17.6953\\
    3.4228&    1.9494&    1.0108&    0.0729&   13.8590\\
    3.4901&    0.3035&    0.3274&    0.0785&    4.1694\\
    3.5576&    0.3257&    0.3797&    0.0661&    5.7405\\
    3.6258&    0.4718&    0.3734&    0.0564&    6.6184\\
    3.7035&    0.2078&    0.8235&    0.0602&   13.6899\\
    3.7845&    0.2703&    0.4291&    0.0587&    7.3109\\
    3.8652&    1.4744&    1.0290&    0.0617&   16.6830\\
    3.9457&    0.1677&    0.5896&    0.0647&    9.1174\\


\enddata

\label{table:irtf_data}
\end{deluxetable*}

\subsection{Palomar/TripleSpec} \label{section:palomar}
\begin{figure}
\centering
\includegraphics[width=1\columnwidth]{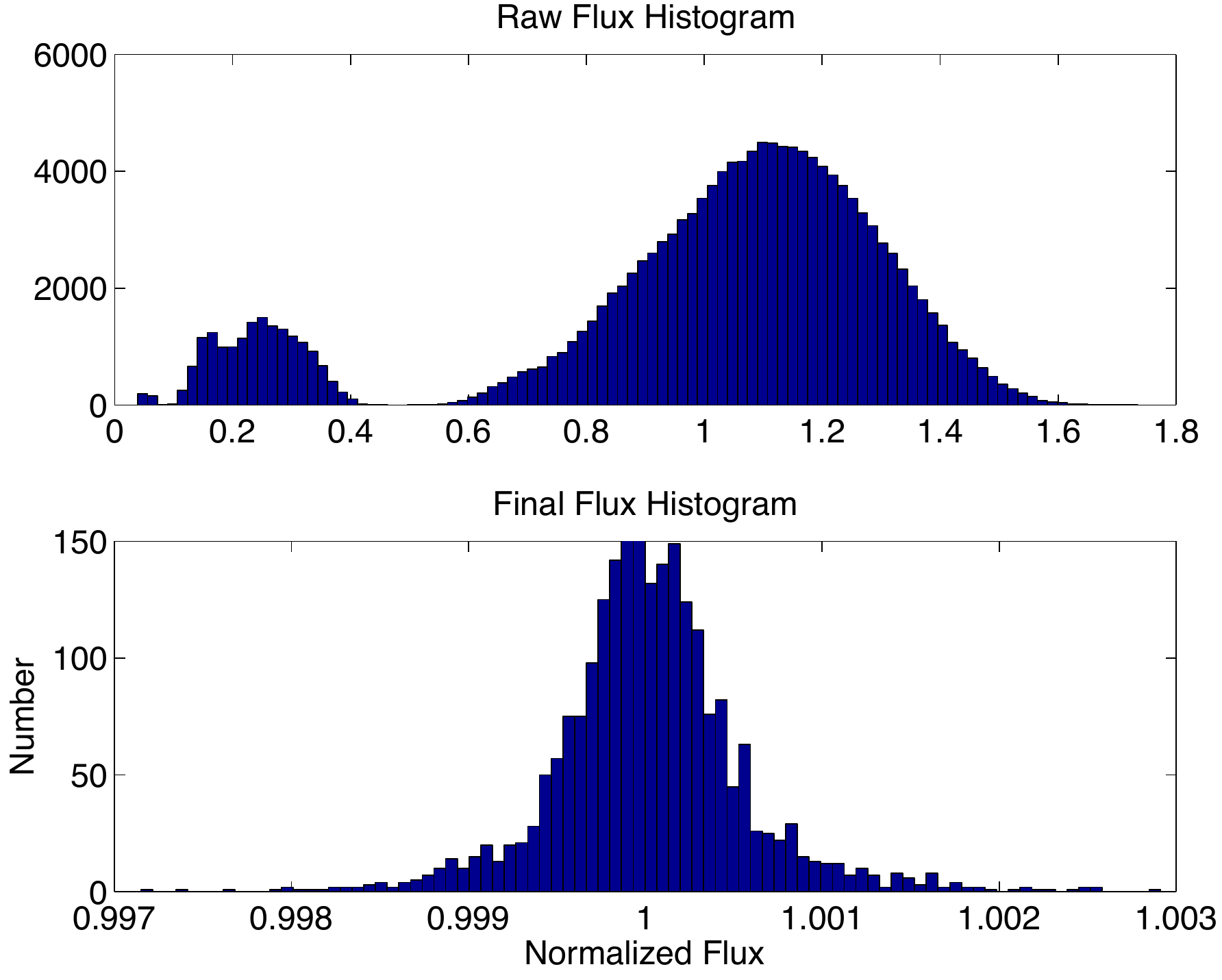}
\caption{\textit{Top:} Histogram of the normalized raw H and K-band Palomar/TripleSpec flux. Note not only the large variance but also the non-Gaussian distribution of the fluxes. The secondary population centered on a flux of $\sim$0.25 corresponds to the autoguider failing $\sim$2.5 hours post-eclipse. \textit{Bottom:} Histogram of the normalized H and K-band flux at the end of the entire reduction. A more Gaussian-like distribution is restored, suggesting the removal of non-Gaussian red noise, and the overall variance is reduced by 2.8 orders of magnitude.}
\label{fig:histogram_palomar}
\end{figure}

\begin{figure}
\centering
\includegraphics[width=1\columnwidth]{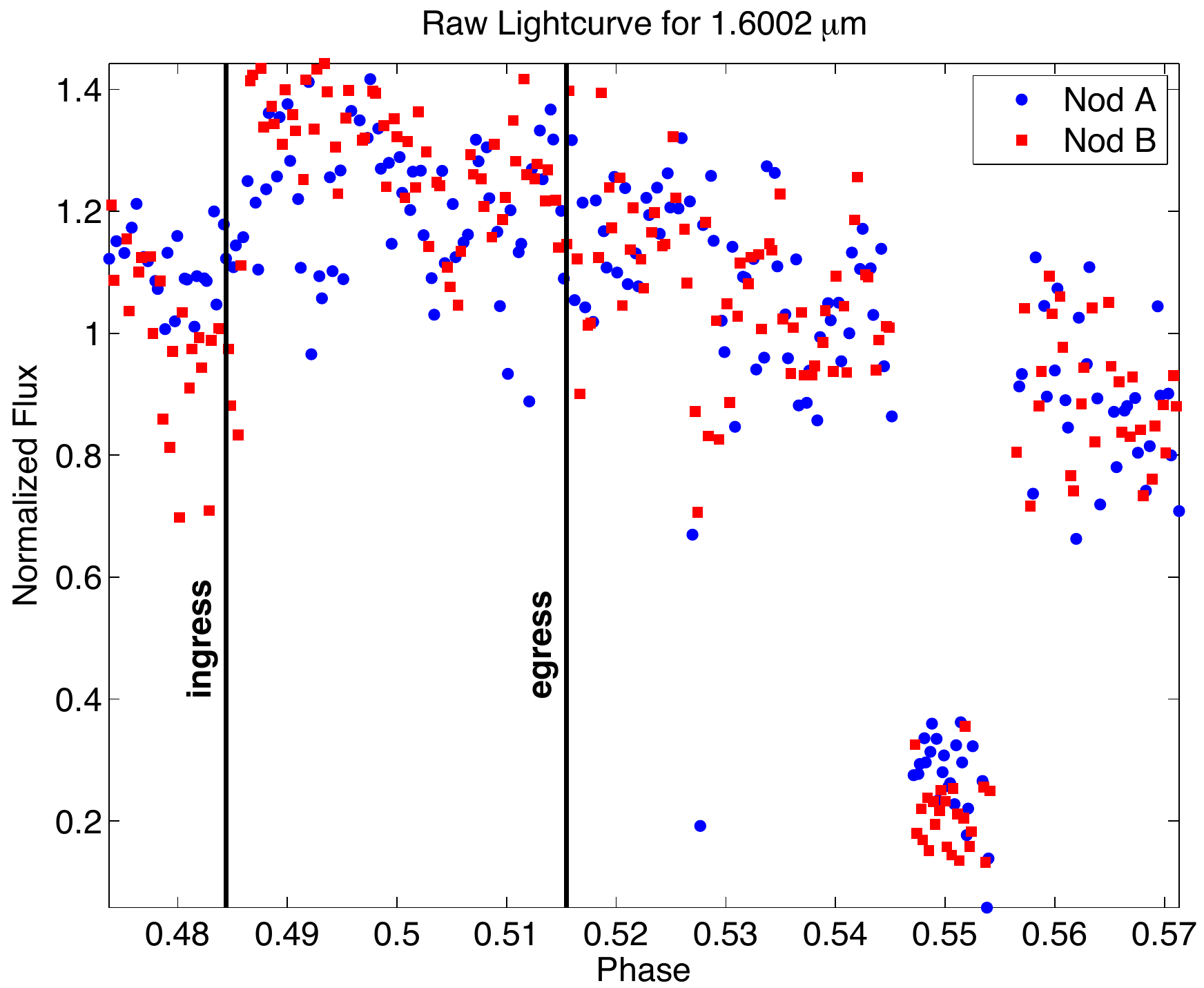}
\caption{A sample raw, normalized Palomar/TripleSpec lightcurve extracted from one wavelength channel (1.6 $\micron$) out of 396 total wavelength channels. At this point, systematic errors cause a variance of $\sim$10--20\%, preventing the detection of the smaller ($\le$0.2\%) eclipse signal. Nod A (blue circles) and nod B (red squares) are plagued by their own distinct set of systematic errors, preventing their combination. We therefore use a principal component analysis to reduce the systematic errors by 2.8 orders of magnitude (Figure~\ref{fig:histogram_palomar}), allowing us to then combine the nods. The drop in flux at a phase of 0.55 ($\sim$2.5 hours post-eclipse) is due to autoguider failure.}
\label{fig:raw_palomar}
\end{figure}

We recorded HD~209458b's 2010 September 4 (UT) secondary eclipse in the near-IR J, H, and K-bands with Palomar/TripleSpec. We observed HD~209458 for $\sim$4.5 hours with an ABBA nodding sequence, taking 396 exposures of 20 or 23 seconds and coadding 2 frames. The night was not optimal: the seeing was at best 1.2'' and at worst 1.5'', causing some ($\lesssim$22\%) of the target to be cut off by the 1'' wide slit and induce variance in the dataset, as indicated in the raw flux histogram (Figure~\ref{fig:histogram_palomar} \textit{Top}) and raw lightcurves (Figure~\ref{fig:raw_palomar}).



Standard methods are adopted to derive the initial lightcurve (Figure~\ref{fig:raw_palomar}). The raw image is first dark-subtracted, flat-fielded, and normalized to an exposure time of 1 second. TripleSpec routines \citep{muirhead11} wavelength-calibrate each image and extracte each spectral order. A Gaussian is fit across the dispersion axis to determine the average horizontal location of the spectral order on the detector and its associated full-width-half-maximum (FWHM). The flux per wavelength channel includes the sum of the pixels in a column along the dispersion axis within 3$\times$FWHM of the order's mean horizontal position. The five spectral orders and the two nods are initially kept separate; systematics are eliminated from each nod and order before they are combined together.

To minimize telluric absorption lines from propagating errors throughout the data, we select wavelengths far from the band edges, i.e. 1.55--1.75 $\micron$ for the H-band and 2.1--2.46 $\micron$ for the K-band, which correspond to 500 and 400 spectral channels, respectively. Due to high telluric absorption at the band edges, the J-band does not have enough spectral channels at relatively higher transmission to resolve the eclipse signal.


The eclipse signal of HD~209458b is not immediately discernible because of the variance from systematic errors. To reduce intra-spectral variations, following \citet{swain10}, each spectrum is normalized by its mean number of counts over all channels (Equations \ref{eqn:normalize} and \ref{eqn:mean}). Each spectrum indicates a slight wavelength shift on the CCD which is corrected by identifying a relatively sharp and bright spectral feature, fitting it with a 4$^{th}$ degree polynomial, and moving it to match up with a mode reference position. Anomalies, such as cosmic rays, are removed with a two-dimensional Wiener filter.



The spectra in each nod are still plagued by separate, distinct errors which prevent their combination. Previous tests of combining the nods and proceeding with the \citet{swain10} and \citet{waldmann12} reductions (normalizing the spectra, binning in Fourier-space, etc.; see Section \ref{section:irtf}) without removing these systematic errors result in error bars three times larger than their current size, preventing the sensitivity necessary to confirm that the Palomar/TripleSpec data agrees with previous \textit{Hubble}/NICMOS measurements. To reduce these errors a principal component analysis (PCA) is conducted to calculate a new set of orthogonal basis vectors that minimizes the amount of variance in a dataset. {Along with PCA, other non-parametric de-trending techniques have been used in the literature, most notably Gaussian Processes, GP \citep[e.g.,][]{gibson12}, and Independent Component Analysis, ICA \citep[e.g.,][]{waldmann12b}. These techniques have been demonstrated for space-based data but not ground-based observations. In the case of our Palomar data, the aim (performed by the PCA analysis) is to reduce the effect of telluric noise which manifests itself in high variance scatter of the raw lightcurve. In the case of GP, the large initial variance makes the model convergence over the modelÕs hyper-parameters difficult. Furthermore, systematic noise not pertaining to the instrument (such as atmospheric seeing and clouds) cannot be captures by a regression of auxiliary optical state vectors as described in \citet{gibson12}. Similarly, the efficiency of an ICA is significantly reduced in the presence of a significant Gaussian noise component in the data. With PCA being a pre-processing step to ICA, the independent components converge to the more classical principal components in high variance data sets \citep[see][]{waldmann12b, waldmann13, waldmann14}. We hence opt for the PCA approach which was demonstrated by \citet{thatte10} to be a robust non-parametric de-trending approach to ground-based data sets.} The PCA analysis produces a new set of components where the first principal component corresponds to a new basis vector containing the most variance, the second principal component contains the second most variance, and so on. Assuming that the eclipse signal is much smaller than sources of error (typically, systematic errors account for $\sim$10--20\% variance while the eclipse signal is only $\le$0.2\%), the first principal component should contain most of the systematic errors. The eclipse signal would then be present in higher order components. Our PCA differs from that used by \citet{thatte10} to recover the secondary eclipse of HD~189733b: while they implement PCA to extract the common eclipse signal from multiple wavelength channels, we use a PCA to find and remove systematic errors common among wavelength channels to clean the data and uncover the underlying eclipse signal.


However PCAs have two caveats: (1) while errors are likely to be contained in the first principal component, they can leak into higher orders, and (2) it is likely that multiple components contain the eclipse signal. Oftentimes the signal and noise are mixed across components as they can have similar amplitudes. We observe a secondary eclipse of an exoplanet on a night with higher seeing and a passing cloud, resulting in large scatter which dwarfs the much smaller eclipse signal. Therefore we assume that the principal component contains most of the noise due to systematic errors and none of the eclipse signal. Since the eclipse signal could be mistakenly removed by eliminating too many components in an attempt to remove the systematic error, only the first principal component is removed. However systematic errors likely leak into the higher ($\ge$2) orders, which we keep, requiring further data reduction in order to resolve the eclipse signal.


The time-varying flux data is prepared for the PCA by first grouping each nod's data into bins of 5 spectral channels. While other bin sizes were tested, this one best-reduces systematic errors as it is large enough to find commonly-shared systematic errors but small enough so that the eclipse signal is not resolved and inadvertently removed by the PCA. Thus, a PCA on this bin size will find a first principal component that contains systematic error common among the spectral channels in the bin, rather than a common eclipse signal. We then run a PCA on the time-varying flux of each bin and eliminated the first principal component to get rid of $\sim$80\% of the wavelength- and time-correlated systematic errors, such as airmass. The remaining ($\ge$2$^{nd}$ order) components are mapped back to obtain cleaned lightcurves. With most of the systematic errors removed, the nods are then combined.

At this point, the stacked spectra indicate large flux variations at the phase of 0.526 ($\sim$30 minutes post-eclispe) and the autoguider failing onward of the phase of 0.55 ($\sim$2.5 hours post-eclipse). However, these two sections of data likely influence the PCA; therefore, we manually remove both of these sections and re-run the entire analysis to better reduce the variance. Despite this ``pre-filtering'' of obvious bad data, the eclipse signal is still not resolved. To enhance the low frequency eclipse signal over the noise, we bin the data by taking the geometric mean in Fourier-space (Equation \ref{eqn:fft}).




The lightcurves still indicate some low-order polynomial curvature introduced by the Fourier transform and from residual systematic errors, such as airmass. In order to correct for this curvature, we fit various amounts of baseline (out-of-eclipse points) to a second degree polynomial. This low-order polynomial is chosen to prevent over-fitting the data with higher-order functions. Since the amount of pre-eclipse data was comparably limited, we vary the pre-eclipse baseline up to half its length. The post-eclipse data is then free to vary so that the total number of baseline points (pre+post-eclipse) is always greater than or equal to the number of second to third contact (in-eclipse) points. As a result, we fit each lightcurve with 3575 different amounts of baseline to a second degree polynomial. After each fit, the standard deviation of the baseline points is stored in an array. The optimal baseline length is the one with the smallest standard deviation. Its associated second degree polynomial fit is then interpolated to the in-eclipse points to correct the entire lightcurve's residual curvature. The eclipse depth is then extracted from the lightcurve by fitting a \citet{mandel02} model lightcurve using the physical parameters of R$_{star}$ = 1.146R$_{\sun}$ \citep{brown01}, R$_{planet}$ = 1.38R$_{Jupiter}$, orbital inclination of 86.59$\degr$, semimajor axis of 0.04747 AU \citep{southworth10}, orbital period of 3.524746 days \citep{torres08}, and period epoch of 2455216.405640 BJD\_UTC  \citep{zellem14}.

\begin{figure}
\centering
\includegraphics[width=1\columnwidth]{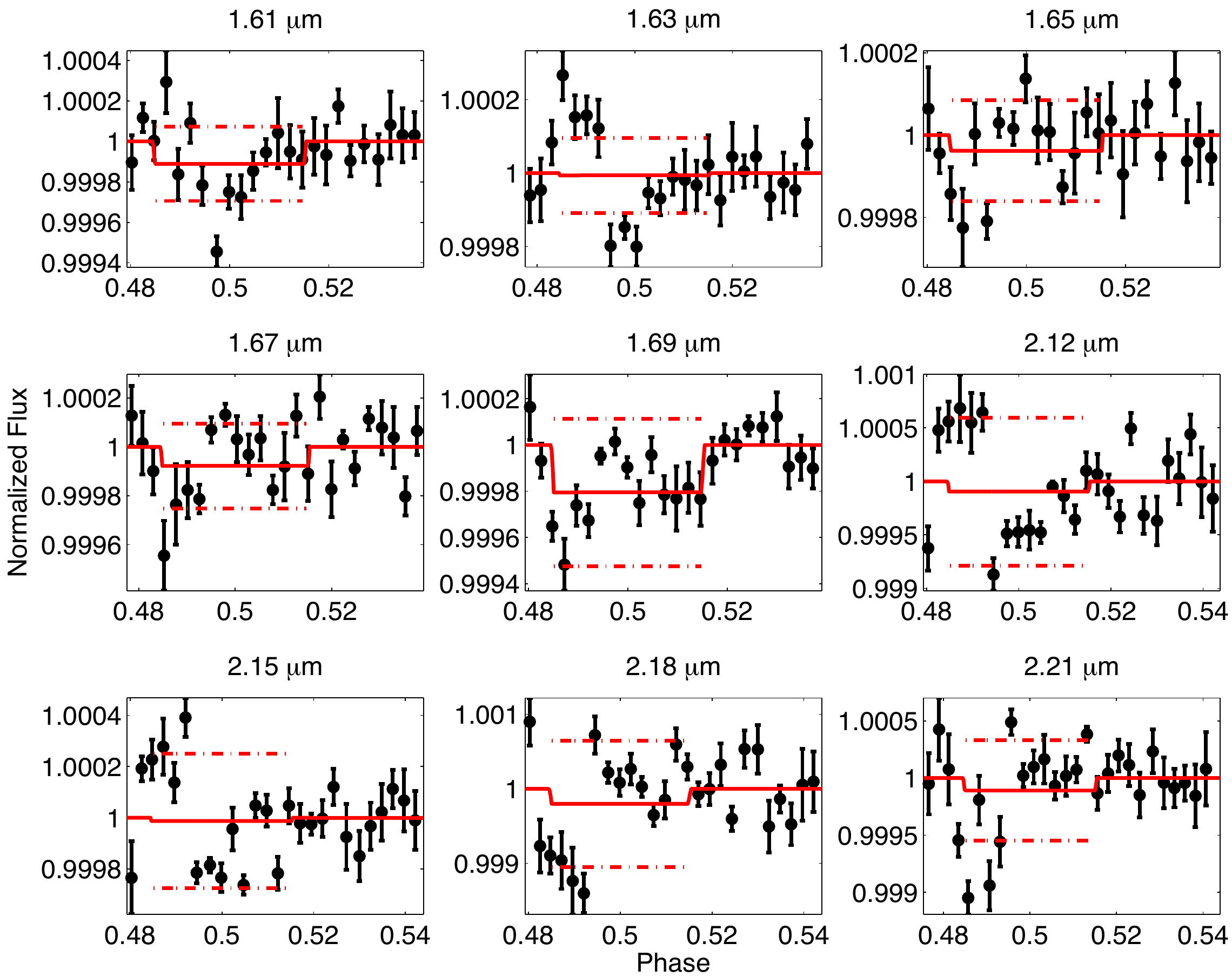}
\caption{The reduced Palomar/TripleSpec lightcurves (both nods combined), with the data binned by 25 datapoints in time, indicate strong upper limits on HD~209458b's emission despite an inconclusive detection of the secondary eclipse due to its low emission (see Figure~\ref{fig:spectrum}). The red line is a model lightcurve fit to the data. The dashed line indicates the 1$\sigma$ uncertainty in our eclipse depths. }
\label{fig:lcs_palomar}
\end{figure}

A bootstrap Monte Carlo simulation estimates the uncertainty of the eclipse depth due to the baseline fitting for each lightcurve. We simulate over 3.5 million lightcurves per original lightcurve, generating over 3.5 million eclipse depths. The standard deviation of all of these simulated eclipse depths reflects the uncertainty of the eclipse depth and the planet-to-star flux ratio. The total eclipse depth uncertainty for each lightcurve is calculated from the root-sum-square of the standard deviations of the in- and out-of-eclipse data points as well as the error introduced from the baseline fit (estimated with a Monte Carlo simulation). Our final, reduced lightcurves are shown in Figure ~\ref{fig:lcs_palomar}.

\begin{figure}
\centering
\includegraphics[width=1\columnwidth]{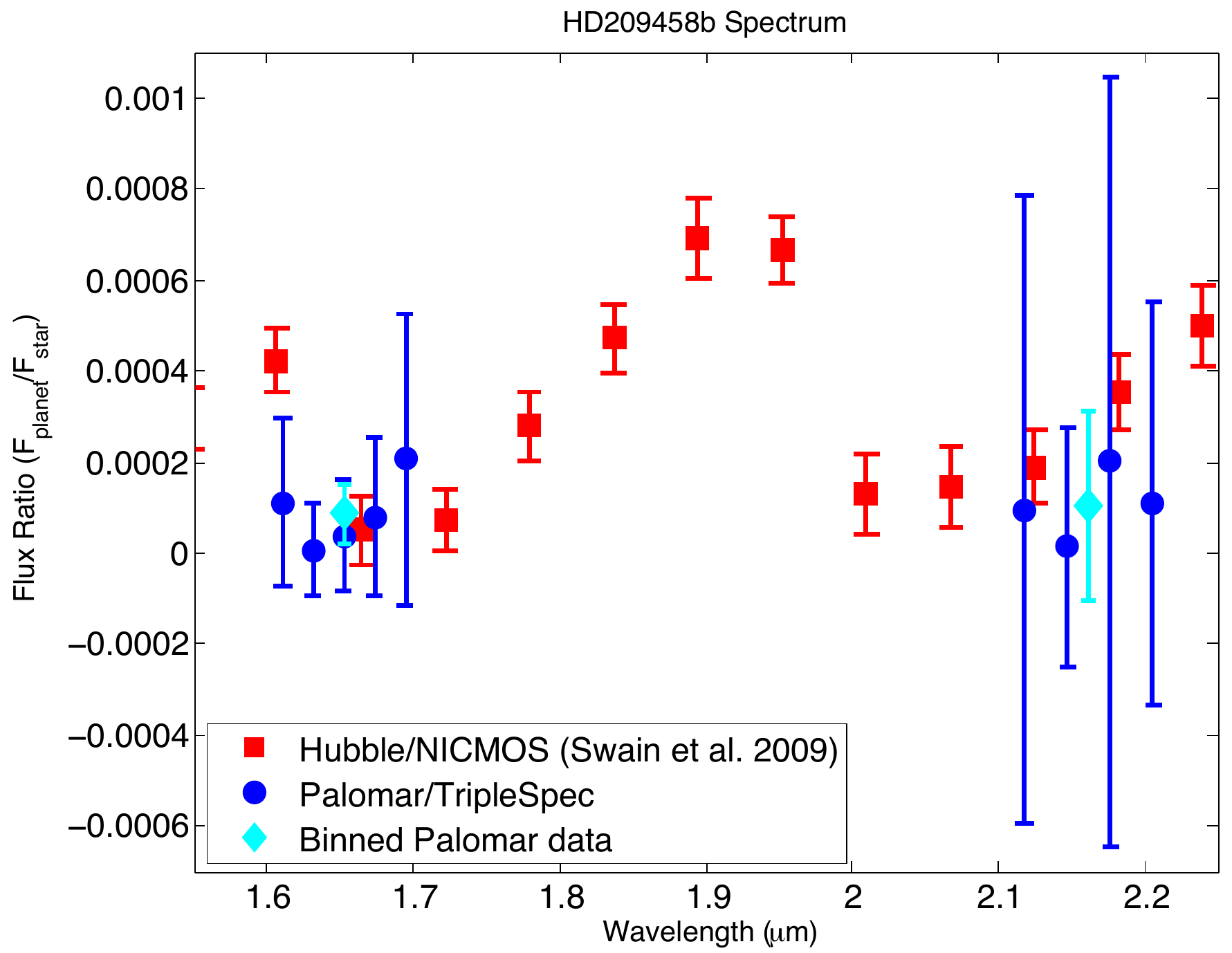}
\caption{The final Palomar/TripleSpec spectrum (blue circles) compared to the \textit{Hubble}/NICMOS spectrum taken by \citet{swain09a} (red squares); please note that we use here a {statistically} independent subset of the published \citet{swain09a} data. The break in our spectrum is due to strong telluric absorption between $\sim$1.8 and $\sim$2 $\micron$. The light blue diamonds are the bins of our H- and K-band spectral points. While all of our unbinned data (blue cirlces) are consistent with a null-detection due to HD~209458b's low emission and residual errors, most of these points agree with \textit{Hubble}/NICMOS to 1$\sigma$ and put strong boundaries on the low flux emission from HD~209458b. These results suggest that this analysis is a viable approach to reducing noisy ground-based data.}
\label{fig:spectrum}
\end{figure}

\begin{deluxetable*}{c c c c c}
\tablewidth{0pt}
\tablecaption{Palomar/TripleSpec Measurements of HD~209458b's Emission}
\tablehead{
\colhead{Wavelength} & \colhead{F$_{\text{planet}}$/F$_{\text{star}}$} & \colhead{Uncertainty} & \colhead{Photon Noise} & \colhead{$\sigma/\sigma_{photon}$}\\
\colhead{($\micron$)} & \colhead{($\times$10$^{3}$)} & \colhead{$\sigma$ ($\times$10$^{3}$)} & \colhead{$\sigma_{photon}$ ($\times$10$^{3}$)} & \colhead{}}
\startdata

    1.6106&    0.1108&    0.1835&    0.0119&   15.4531\\
    1.6314&    0.0064&    0.1023&    0.0116&    8.8418\\
    1.6526&    0.0380&    0.1224&    0.0114&   10.7071\\
    1.6734&    0.0782&    0.1740&    0.0115&   15.1614\\
    1.6943&    0.2053&    0.3193&    0.0116&   27.6098\\
    2.1182&    0.0952&    0.6897&    0.0111&   62.0932\\
    2.1478&    0.0131&    0.2626&    0.0109&   23.9908\\
    2.1771&    0.2002&    0.8475&    0.0111&   76.4640\\
    2.2058&    0.1088&    0.4409&    0.0111&   39.7775\\

\enddata

\label{table:palomar_data}
\end{deluxetable*}


\section{Discussion}
\subsection{Results}
Here we present the IRTF/SpeX and Palomar/TripleSpec H, K, and L-band emission spectra of the hot Jupiter HD~209458b measured from two secondary eclipses. Systematic errors, such as telescope jitter and airmass variations, can cause data scatter on the order of $\sim$10-20\% which dwarf HD~209458b's secondary eclipse depth ($\sim$0.2\%). These common-mode errors are removed with a normalization routine for IRTF/SpeX (Section \ref{section:irtf}) and a principal component analysis for Palomar/TripleSpec (Section \ref{section:palomar}). The low frequency eclipse signal is enhanced over the remaining noise by binning spectral channels in Fourier-space. The IRTF/SpeX reduction decreases the variance by 1.6 orders of magnitude (Figure~\ref{fig:fluxhist_irtf}) to get within 2.9 times the photon noise limit (Table~\ref{table:irtf_data}). Due to high telluric absorption at the band edges, the Palomar/TripleSpec J-band lacks a sufficient number of spectral channels to resolve the eclipse signal. Yet for the H and K-bands, the Palomar/TripleSpec reduction decreases the variance by 2.8 orders of magnitude (Figure~\ref{fig:histogram_palomar}) to get within 8.8 times the photon noise limit (Table~\ref{table:palomar_data}). Both reductions restore a more symmetrical shape in the final flux histograms, suggesting a reduction of non-Gaussian red noise (Figures~\ref{fig:fluxhist_irtf} and~\ref{fig:histogram_palomar}). Despite the persistence of red noise, as suggested by the non-Gaussian shape of the final flux histograms (Figures~\ref{fig:fluxhist_irtf} and~\ref{fig:histogram_palomar}), we reduce the IRTF/SpeX K-band mean variance to 175 ppm and the Palomar/TripleSpec binned K-band uncertainty to 208 ppm. For comparison, \citet{swain10} found a mean K-band variance of $\sim$150 ppm while observing the secondary eclipse of HD~189733b (V-mag = 7.68, K-mag = 5.54) with IRTF/SpeX. After scaling our IRTF/SpeX and Palomar/TripleSpec binned K-band uncertainties while accounting for the difference in brightnesses between HD~209458b and HD~189733b and primary mirror diameters of IRTF (3~meters) and Palomar (5.08~meters), we achieve a similar variance (123 ppm and 174 ppm, respectively) to \citet{swain10}.  However the Hale telescope, due to its larger primary mirror (5.08 m), can theoretically achieve a variance $\sim$60\% smaller than IRTF on the same targets under similar observation conditions. We do not achieve this theoretical precision for Palomar likely due to the IRTF's superior observing conditions on Mauna Kea. In the IRTF/SpeX data, we detect non-zero eclipse depths in three of the nine K-band lightcurves (Figure~\ref{fig:finallcs_irtf_k}) and in seven of the fourteen L-band lightcurves (Figure~\ref{fig:finallcs_irtf_l}). All of the Palomar/TripleSpec lightcurves are consistent with a null detection (Figure~\ref{fig:lcs_palomar}). Both datasets, despite the different platforms and reduction schemes, agree both with each other and with previous \textit{Hubble}/NICMOS \citep{swain09a} and $Spitzer$/IRAC1 \citep{knutson08} observations to 1$\sigma$ {and a new $Spitzer$/IRAC1 measurement \citep{diamondlowe14} to 1.5$\sigma$} (Figures~\ref{fig:spectrum_irtf_k}, \ref{fig:spectrum_irtf_l}, and~\ref{fig:spectrum}), suggesting the validity of the observation and reduction techniques employed here. In addition, the K and L-band measurements resemble previous IRTF/SpeX HD~209458b emission contrast spectra \citep{richardson03b}.

\subsection{Palomar/TripleSpec's Ability to Measure Other Targets}
The HD~209458b binned K-band uncertainty (208 ppm) suggests that Palomar/TripleSpec is capable of high-precision exoplanet measurements of both primary transits and secondary eclipses. In particular, the number of secondary eclipse higher SNR targets for Palomar/TripleSpec is estimated from the host star's radius R$_{s}$ and effective temperature T$_{s}$, and the planet's semimajor axis $a$. Assuming an albedo of 0 and efficient energy transport to the night side, the planet's equilibrium temperature T$_{p}$ is:
\begin{eqnarray}
T_{p} = T_{s} \sqrt{\frac{R_{s}}{2a}}.
\end{eqnarray}
Note that this simple estimate underestimates the dayside temperature of hotter planets (T$_{irradiated}$ $\ga$ 2000 K), which are predicted to redistribute energy from the dayside to the nightside less efficiently \citep[e.g.,][]{cowan11, perna12, perez-becker13}. Therefore our estimated SNRs of the hotter exoplanets are very conservative. Assuming that the host star and planet emit as black bodies, we estimate the planet-to-star flux ratio F$_{p}$/F$_{s}$ at the center of the K$_{s}$ band (2.15 $\micron$). Note that while planetary spectra diverge from blackbody emission, this assumption is necessary as we do not know the planet's composition a-priori. To estimate each target's theoretical uncertainty $\sigma$, we scale HD~209458b's binned K-band uncertainty ($\sigma_{\text{HD\,209}}$  = 208 ppm) to each system using their respective K$_{s}$ magnitudes $m$:
\begin{eqnarray}
\sigma &=& \sigma_{\text{HD\,209}} \sqrt{\frac{F_{\text{HD\,209}}}{F}} \nonumber \\
&=& \sigma_{\text{HD\,209}} \sqrt{10^{-0.4\left(m_{\text{HD\,209}} - m\right)}}
\end{eqnarray}
where $F_{\text{HD\,209}}$ is the flux of HD~209458, $F$ is the flux of the system's host star, $m_{\text{HD\,209}}$ is the K$_{s}$ magnitude of HD~209458 (K$_{s}$-mag = 6.307), and $m$ is the K$_{s}$ magnitude of the system's host star. According to these calculations, for example, Palomar/TripleSpec can measure WASP-33b's K-band emission with a SNR~$\approx4$ (Table~\ref{table:emission}).


\begin{deluxetable}{l c c c c}
\tablewidth{0pt}
\tablecaption{Potential exoplanet targets for measuring the secondary eclipse emission with Palomar/TripleSpec.}
\tablehead{
\colhead{Exoplanet} & \colhead{K$_{s}$-mag} & \colhead{F$_{planet}$/F$_{star}$\tablenotemark{a}} & \colhead{Uncertainty\tablenotemark{b}} & \colhead{SNR}\\
\colhead{} & \colhead{} & \colhead{($\times$10$^{-3}$)} & \colhead{($\times$10$^{-3}$)} & \colhead{}}
\startdata



WASP-33 b &       7.46800 &       1.44271 &      0.354800 &       4.06625 \\
HD 189733 b &       5.54100 &      0.241288 &      0.146077 &       1.65178 \\
WASP-12 b &       10.1880 &       1.86983 &       1.24160 &       1.50598 \\
WASP-77 A b &       8.40500 &      0.716147 &      0.546240 &       1.31105 \\
KOI-13 b &       9.42500 &       1.03594 &      0.873742 &       1.18563 \\
HAT-P-32 b &       9.99000 &       1.20003 &       1.13340 &       1.05878 \\
WASP-103 b &       10.7670 &       1.69008 &       1.62100 &       1.04262 \\

\enddata

\tablenotetext{a}{estimated flux ratio of the planet to the star (see text)}
\tablenotetext{b}{scaled from the binned K-band uncertainty measured here for HD~209458b (208 ppm) to each system using their respective K$_{s}$ magnitudes}
\label{table:emission}
\end{deluxetable}

The theoretical Palomar/TripleSpec K-band SNRs of primary transit measurements are calculated by first estimating the scale height $H$ of each planet, assuming a mean atmospheric temperature equal to the planet's equilibrium temperature and a H$_{2}$-dominated atmosphere. If the planet's atmosphere is optically thick at 5 scale heights in the K$_{s}$ band, then the primary transit depth is:
\begin{eqnarray}
\text{transit depth} = \left( \frac{R_{p}+5H} {R_{s}} \right) ^{2}.
\end{eqnarray}
{Since atmospheres can change drastically between different targets, we avoid adding any additional interpretation or assumptions into their composition or structure with a ``channel to channel'' SNR calculation. We instead calculate a ``broad band'' SNR across the entire K$_{s}$ band and avoid giving SNRs on the atmospheric absorption as doing so presupposes knowledge of the planet's atmospheric composition. The typical channel-to-channel signal from a spectrum will be approximately a factor 10--100 smaller than the total photometric $K_{s}$-band transit depth calculated in Table~\ref{table:absorption}.} Palomar/TripleSpec can theoretically measure the primary transit absorption of 14 exoplanets with SNR $\ge$ 20 and 107 additional exoplanets with SNR $\ge$ 3 (Table~\ref{table:absorption}).

\begin{deluxetable*}{l c c c c c}
\tablewidth{0pt}
\tablecaption{Potential exoplanet targets for measuring the primary transit with Palomar/TripleSpec.}
\tablehead{
\colhead{Exoplanet} & \colhead{K$_{s}$-mag} & \colhead{H\tablenotemark{a}} & \colhead{Absorption\tablenotemark{b}} & \colhead{Uncertainty\tablenotemark{c}} & \colhead{SNR\tablenotemark{d}} \\ 
\colhead{}  & \colhead{} & \colhead{(km)} & \colhead{($\times$10$^{-3}$)} & \colhead{($\times$10$^{-3}$)} & \colhead{}
}

\startdata

HD 189733 b &       5.54100 &       217.994 &       23.5265 &      0.146077 &       161.055 \\
WASP-80 b &       8.35100 &       214.307 &       28.9982 &      0.532823 &       54.4237 \\
GJ 436 b &       6.07300 &       203.274 &       7.18361 &      0.186631 &       38.4910 \\
HD 80606 b &       7.31500 &       17.3602 &       11.1666 &      0.330661 &       33.7707 \\
WASP-43 b &       9.26700 &       107.398 &       24.8430 &      0.812425 &       30.5789 \\
WASP-77 A b &       8.40500 &       223.018 &       16.6508 &      0.546240 &       30.4826 \\
WASP-34 b &       8.79200 &       475.315 &       18.3705 &      0.652805 &       28.1409 \\
HAT-P-20 b &       8.60100 &       16.0539 &       15.8112 &      0.597838 &       26.4474 \\
HAT-P-22 b &       7.83700 &       111.412 &       11.0576 &      0.420517 &       26.2954 \\
HAT-P-17 b &       8.54400 &       240.431 &       15.2178 &      0.582349 &       26.1318 \\
GJ 1214 b &       8.78200 &       251.885 &       14.9799 &      0.649805 &       23.0529 \\
WASP-10 b &       9.98300 &       56.2980 &       24.3706 &       1.12975 &       21.5716 \\
HAT-P-32 b &       9.99000 &       1285.69 &       23.8054 &       1.13340 &       21.0035 \\
CoRoT-2 b &       10.3100 &       161.761 &       27.1131 &       1.31336 &       20.6441 \\

\enddata
\tablenotetext{a}{scale height $H$ calculated by assuming a H$_{2}$-dominated atmosphere and that the mean atmospheric temperature is equal to the planet's equilibrium temperature}
\tablenotetext{b}{estimated lightcurve depth during primary transit; {assuming that the planet's atmosphere is optically thick at 5 scale heights $H$, we estimate the primary transit depth = $\left( \frac{R_{p}+5H} {R_{s}} \right) ^{2}$; please note that the typical channel-to-channel signal from a spectrum will be approximately a factor 10--100 smaller than the total photometric $K_{s}$-band transit depth calculated here}}
\tablenotetext{c}{scaled from the binned K-band uncertainty measured here for HD~209458b (208 ppm) to each system using their respective K$_{s}$ magnitudes}
\tablenotetext{d}{While Palomar/TripleSpec has the theoretical capability to measure the primary transit absorption of 121 exoplanets to SNR $\ge$ 3, here we only list the 14 exoplanets which have SNR $\ge$ 20 in the interest of space.}
\tablenotetext{}{Table~\ref{table:absorption} is published in its entirety in the electronic edition of \apj. A portion is shown here for guidance regarding its form and content.}
\label{table:absorption}
\end{deluxetable*}

\subsection{HD~209458b's L-band Emission Spectrum}
\begin{figure}
\centering
\includegraphics[width=1\columnwidth]{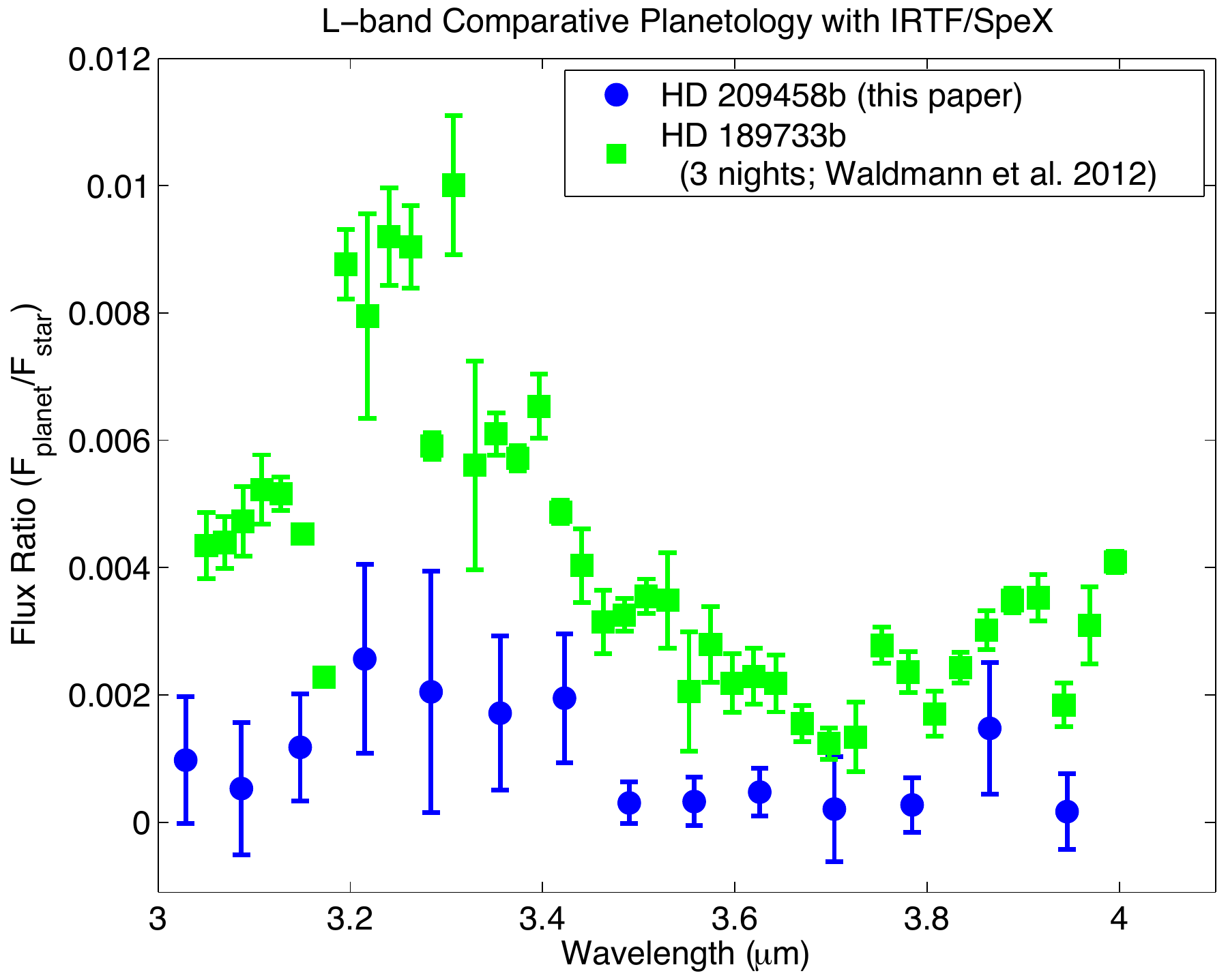}
\caption{Our final IRTF/SpeX HD~209458b L-band emission spectrum (blue circles) versus IRTF/SpeX L-band emission spectra of HD~189733b binned over 3 nights \citep[][green squares]{waldmann12}. HD~189733b exhibits bright emission at 3.3~$\micron$, which is attributed to non-LTE CH$_{4}$ $\nu_{3}$ fluorescence \citep{drossart11}. HD~209458b does not exhibit similar emission to $\sim$3$\sigma$, suggesting that fluorescence is not needed (Figure~\ref{fig:spectrum_rt}).}
\label{fig:spectrum_189vs209}
\end{figure}

The L-band data (Figure~\ref{fig:spectrum_irtf_l}) presented here is the first absolute measurement of any exoplanet other than HD~189733b. The resultant L-band spectrum that we derive for HD~209458b differs significantly from that observed 3 times for HD~189733b (Figure~\ref{fig:spectrum_189vs209}). Unlike HD~189733b, HD~209458b does not exhibit bright emission at 3.3 $\micron$ to $\sim$3$\sigma$, suggesting that HD~189733b's emission is specific to that exoplanet. HD~189733b's bright L-band emission is attributed to non-LTE CH$_{4}$ $\nu_{3}$ fluorescence \citep{swain10,drossart11,waldmann12}. Considering thermochemical equilibrium \citep{moses11}, the relatively cooler HD~189733b (T$_{eff} \approx 1200$ K) is predicted to have a higher CH$_{4}$ abundance than the hotter HD~209458b (T$_{eff} \approx 1450$ K); thus one might expect for HD~209458b to have weaker CH$_{4}$ $\nu_{3}$ emission. The theory that cooler objects have more CH$_{4}$ is supported by brown dwarf studies \citep[e.g.,][]{burrows97,burgasser08} and further reinforced by the lack of detection of a bright 3.3~$\micron$ feature on HD~209458b.



\begin{figure}
\centering
\includegraphics[width=1\columnwidth]{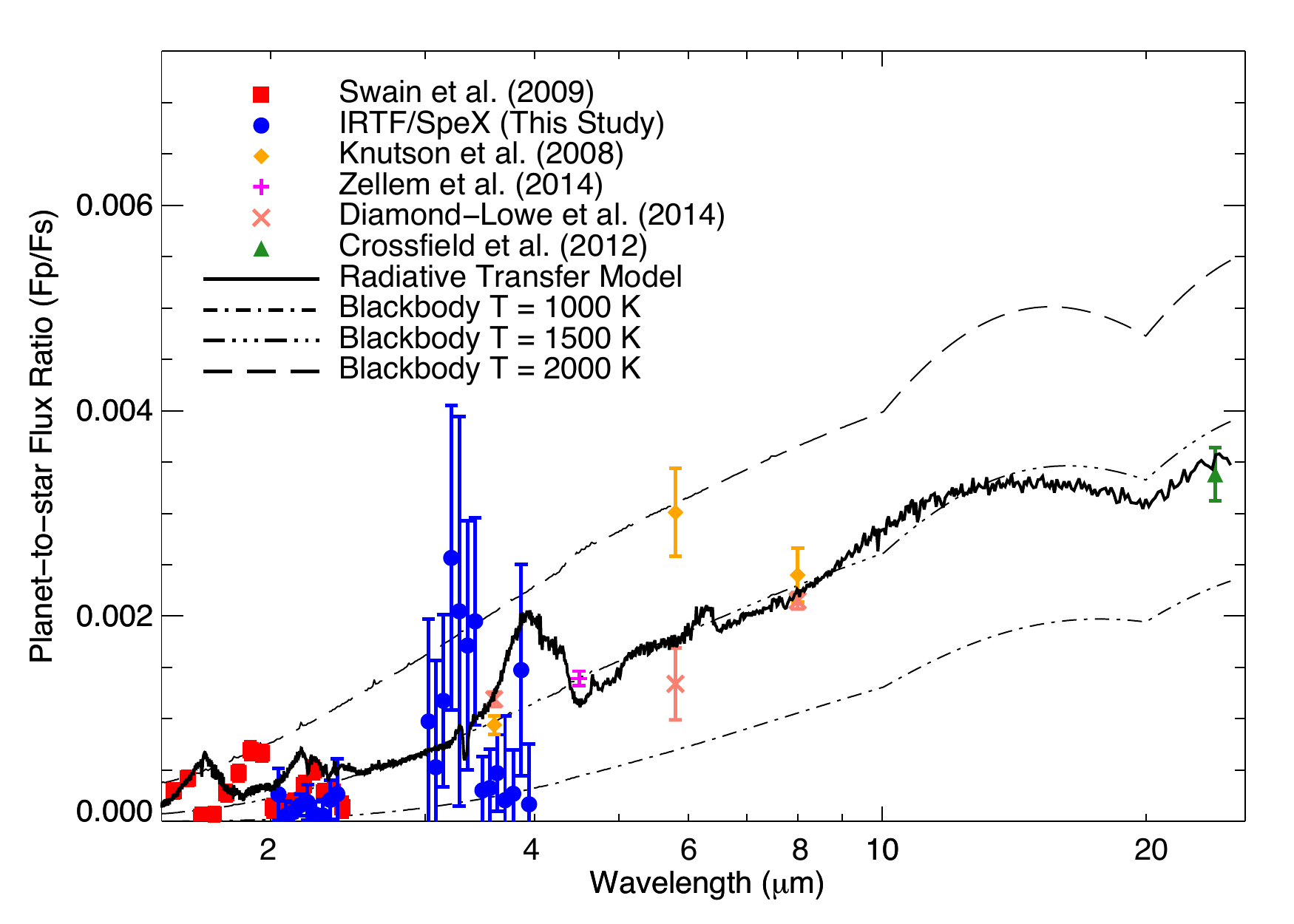}
\caption{A comparison of existent HD~209458b secondary eclipse data {\citep{swain09a, knutson08, zellem14, diamondlowe14, swain08, crossfield12b}} with our IRTF/SpeX K and L-band data (blue circles). A radiative transfer model (black line), whose inputs are based on the \citet{moses11} atmospheric model, fits all of the data nominally well, including our new IRTF/SpeX data. Also included are blackbody emission curves for T~$=1000$~K (dash dot), T~$=1500$~K (dash dot dot dot), and T~$=2000$~K (dash).}
\label{fig:spectrum_rt}
\end{figure}

In order to explore the question of whether the HD~209458b L-band measurements require non-LTE emission, we compare these data with a spectrum of a model atmosphere (Figure~\ref{fig:spectrum_rt}). {Radiative transfer analyses of HD~209458b assume solar abundances and the presence of major sources of opacity expected for
an extrasolar planet with HD~209458b's equilibrium temperature, considering
thermochemical equilibrium and disequilibrium processes \citep{moses11},
i.e. H$_{2}$O, CH$_{4}$, CO  and CO$_{2}$, and using a temperature-pressure profile consistent with current GCM models \citep{showman09}. Using k-coefficients of \citet{griffith14} the emission
of the exoplanet is calculated over pressure levels
ranging from 10 to 10$^{-5}$ bar, assuming LTE and the appropriate stellar
model from \citet{castellikurucz04} for the host
star's flux.} Note however that the model does not use the hot CH$_{4}$ lines of \citet{exomol14}--we leave a full radiative transfer analysis implementing these line lists to a future study. The agreement between the model and the data further suggests that fluorescence is not needed to explain HD~209458b's emission, unlike HD~189733b \citep{swain10,waldmann12}.

%




\section{Conclusions}


IRTF/SpeX, until the launch of the \emph{James Webb Space Telescope}, is one of the few platforms currently capable of measuring an exoplanet's low resolution L-band emission spectrum, which probes the CH$_{4}$ $\nu_{3}$ band. Combined with Palomar/TripleSpec, with simultaneous J, H, and K-band coverage and a large primary mirror, these two platforms have the potential to observe wavelength regions where the major C and O bearing species (H$_{2}$O, CO, CO$_{2}$, and CH$_{4}$) emit. Thus, measurements of these wavelengths coupled with radiative transfer analyses constrain the abundances of these molecules. In addition, these two platforms can help identify targets for future missions, such as the \emph{James Webb Space Telescope}. {IRTF/SpeX and potentially Palomar/TripleSpec, combined with significant data reduction as demonstrated here, are reliable instruments to use for high-precision ground-based eclipsing exoplanet spectroscopy.}

%
%
%

\section{Acknowledgements}
RTZ and CAG are supported by the NASA Planetary Atmospheres Program. Part of the research was carried out at the Jet Propulsion Laboratory, California Institute of Technology, under contract with the National Aeronautics and Space Administration.

The authors would like to thank Ming Zhao for observing HD~209458b with Palomar/TripleSpec.

RTZ would like to thank Ian J. M. Crossfield and Michael R. Line for their helpful discussions.

We would like to thank the referee for their helpful comments and suggestions.

\bibliography{references}

\end{document}